\documentclass{article}

\usepackage{amsmath,amssymb}
\usepackage{graphicx}
\begin{document}


\title{Enhancement of non-resonant dielectric cloaks using anisotropic composites}


\author{Akihiro Takezawa}
\author{Akihiro Takezawa
\thanks{Corresponding author, akihiro@hiroshima-u.ac.jp} \thanks{Division of Mechanical Systems and Applied Mechanics, Institute of Engineering, Hiroshima University, 1-4-1 Kagamiyama, Higashi-Hiroshima, Hiroshima, 739-8527, Japan}
\and
Mitsuru Kitamura$^*$
}

\date{\today}

\maketitle

\begin{abstract}
Cloaking techniques conceal objects by controlling the flow of electromagnetic waves to minimize scattering. Herein, the effectiveness of homogenized anisotropic materials in non-resonant dielectric multilayer cloaking is studied. Because existing multilayer cloaking by isotropic materials can be regarded as homogenous anisotropic cloaking from a macroscopic view, anisotropic materials can be efficiently designed through optimization of their physical properties. Anisotropic properties can be realized in two-phase composites if the physical properties of the material are within appropriate bounds. The optimized anisotropic physical properties are identified by a numerical optimization technique based on a full-wave simulation using the finite element method. The cloaking performance measured by the total scattering width is improved by about 2.8\% and 25\% in eight- and three-layer cylindrical cloaking materials, respectively, compared with multilayer cloaking by isotropic materials. In all cloaking examples, the optimized microstructures of the two-phase composites are identified as the simple lamination of two materials, which maximizes the anisotropy. The same performance as published for eight-layer cloaking by isotropic materials is achieved by three-layer cloaking using the anisotropic material. Cloaking with an approximately 50\% reduction of total scattering width is achieved even in an octagonal object. Since the cloaking effect can be realized using just a few layers of the laminated anisotropic dielectric composite, this may have an advantage in the mass production of cloaking devices.\end{abstract}


\maketitle

\section{INTRODUCTION}
By controlling the flow of electromagnetic waves to minimize the scattering from an object, the object can be concealed from the wave. This technology is called "cloaking" and is the subject of intense investigation. The representative cloaking theory assumes an inhomogeneous anisotropic material for the cloaking layer, the physical properties of which are determined by the ordinal coordinate transformation \cite{pendry2006,schurig2006,cummer2006,rahm2008}. However, the theoretically obtained physical properties cannot be realized in conventional materials. Because of the requirement for the material to possess both a permittivity and permeability of less than 1, this can only be realized in metamaterials. There have been several experimental and numerical reports that have used metamaterials, for example, as a sprit-ring resonator \cite{pendry2006}, dielectric and metal wire composite \cite{cai2007}, and cylindrical dielectric \cite{gaillot2008,cai2008}. 

Another way of cloaking is the cancellation of scattering using plasmonic materials \cite{alu2008,alu2009}. However, since these materials utilize local resonance inside the cloaking region, the propagating wave suffers attenuation due to energy loss. In the case of visible right, this leads to a darkening of the object. To resolve this issue and to achieve easy implementation, dielectric non-resonant cloaking has been studied \cite{andkjaer2011,wang2013}. A recent report \cite{wang2013} shows that non-resonant cloaking can be achieved by multilayered dielectrics. In this type of cloaking, increasing the number of layers can be effective for improving the performance. However, this also increases the number of different kinds of dielectric used, which leads to manufacturing complexity.

Numerical optimization techniques are a powerful tool for optimizing the physical properties of materials in the search for cloaking devices \cite{popa2009, andkjaer2011, yu2011, okada2012, song2013}. In the case of isotropic dielectric multilayer cloaking\cite{wang2013}, the physical properties of each layer are determined by an optimization algorithm.

Multilayers have been introduced to achieve anisotropy in the cloaking region by using only isotropic materials. From the macroscopic view, the layered isotropic material can be regarded as a homogenous anisotropic material \cite{milton2002}. This approach therefore focuses on the microscopic structure to achieve macroscopic anisotropy. In contrast, we have taken the opposite approach in the present work to study non-resonant dielectric cloaking. The macroscopic anisotropic physical properties were first studied by numerical optimization and the microscopic structures required to achieve them were analyzed. The composites forming each layer were assumed to be made by two kinds of isotropic material. If the resulting optimized physical properties fall within the range of possible properties the material\cite{milton2002}, the material can be implemented as an actual composite. Since one layer composed of anisotropic material essentially functioned as a multilayer composed of isotropic materials, effective non-resonant dielectric cloaking is expected to be achieved using fewer layers than in previous studies.

\section{FORMULATION}
\subsection{Problem statement}
\begin{figure}[htbp]
\centering
\includegraphics[scale=1.0,clip]{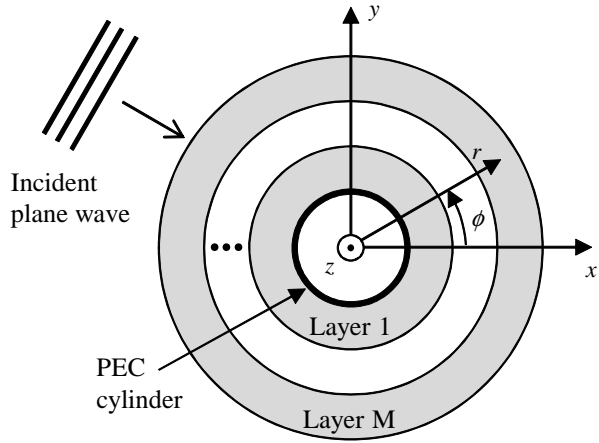}
\caption{Outline of multilayer dielectric cloaking. The wave propagation is analyzed in the $x-y$ plane. The anisotropic material properties are set according to the $r-\phi$ coordinate system.}
\label{model}
\end{figure}

Let us consider a 2D cylindrical perfect electric conductor (PEC) object in an open space, cloaked by the multilayer dielectric as shown in Fig. \ref{model}. The dielectric layers are composed of ordinary homogenous isotropic or anisotropic materials. Steady-state full-wave propagation is analyzed in the domain including the cloaking object. To perform the simulation on an open space, the domain is surrounded by a perfectly matched layer domain. We evaluate the scattering from the object against the H-Polarized incident wave. The wave propagation in the domain is dominated by the following Maxwell equation in the magnetic form:
\begin{equation}
\nabla\times\left(\boldsymbol{\varepsilon}^{-1}\nabla\times\bold{H}\right) + \frac{\omega^2}{c^2}\boldsymbol{\mu}\bold{H} = 0
\label{eq01}
\end{equation},
where $\bold{H}$, $\boldsymbol{\varepsilon}$, $\boldsymbol{\mu}$ and $c$ are the magnetic field vector, relative permittivity tensor, relative permeability tensor and the speed of light. For an H-polarized incident wave ($\bold{H_0}=(0,0,\text{exp}(-jk_0\bold{x}))^T$ where $\bold{x}$ is wave propagation direction vector),  \eqref{eq01} is solved for the scattering of the magnetic field $\bold{H_{sc}}$ with the form of $\bold{H} = \bold{H_0} + \bold{H_{sc}}$.

The anisotropic dielectric that comprises the cloaking layer is assumed to be homogenous and its principal directions are matched to the $r$ and $\phi$ directions of a cylindrical coordinate system $(r,\phi,z)$ surrounding the cloaking object. The material properties used in the analysis of the H-polarized wave are $\varepsilon_r$ and $\varepsilon_\theta$. The material properties in the ordinal coordinate system are calculated by a coordinate transformation.

Assuming that the scattering is evaluated far enough from the object, the cloaking performance is evaluated by the far-field value of the scattered magnetic field. The far magnetic field is calculated from the near-field according to the Stratton--Chu formula \cite{jin2009}. The boundary integral is performed on a 2$\lambda$ radius circle, the center of which coincides with the center of the cloaked object. The cloaking performance is evaluated using the scattering width, i.e. the radar cross section per unit length, which is calculated using the far magnetic field $\bold{H_{scfar}}$:
\begin{equation}
\sigma(\phi) = 2\pi r_0\frac{|\bold{H_{scfar}}|^2}{|\bold{H_0}|^2}
\label{eq03}
\end{equation},
where $r_0$ is the distance from the object to the evaluation point.

The physical properties of the cloaking layer that minimize the scattering width are determined by numerical optimization based on a full-wave simulation by the finite element method. The $r$ and $\phi$ elements of the relative permittivity are regarded as design variables, which are varied by a gradient approach.

\subsection{Bounds of the composite's physical properties}
\begin{figure}[htbp]
\centering
\includegraphics[scale=1.0,clip]{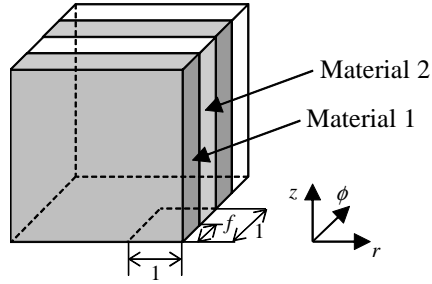}
\caption{Outline of the lamination pattern corresponding to the Wiener upper and lower bounds. This lamination corresponds to the state where $\varepsilon_r$ and $\varepsilon_\phi$ are on the upper and lower bounds, respectively, with a volume fraction $f$ for material 1.}
\label{laminate}
\end{figure}

For the anisotropic material used in this study, a composite composed of two isotropic materials is considered. In the design of the composite, the bounds of its physical properties must be defined. Wiener arithmetic and harmonic bounds \cite{milton2002} are introduced, which are formulated as follows:
\begin{equation}
\left(\frac{f}{\varepsilon_1} + \frac{1-f}{\varepsilon_2} \right) \le \varepsilon_r, \varepsilon_\phi\ \le f\varepsilon_1 + (1-f)\varepsilon_2
\label{eq04}
\end{equation},
where $\varepsilon_1$ and $\varepsilon_2$ are the relative permittivities of material 1 and 2, which comprise the composite, and $f$ is the volume fraction of material 1, which has a range $0\le f\le1$. The lower and upper bounds correspond to the effective permittivity of the composite with a laminated unit cell, the laminated directions of which are parallel and perpendicular to the evaluation direction as shown in Fig. \ref{laminate}. Under the assumption that $\varepsilon_r \le \varepsilon_\phi$ and that the value of $\varepsilon_r$ is pre-defined, the volume fraction of material 1 takes the range $(\varepsilon_r - \varepsilon_2)/(\varepsilon_1 - \varepsilon_2)\le f<1$. The lower bound of $f$ corresponds to the lamination pattern shown in Fig. \ref{laminate}. The definite upper bound of $f$ can be derived by introducing a bound for an isotropic composite such as the Hashin-Shtrikman bound \cite{milton2002}, the detailed calculation of which is omitted here. The bound for $\varepsilon_\phi$ is derived as follows using the minimum volume fraction of material 1:
\begin{equation}
\left(\frac{f_{\min}}{\varepsilon_1} + \frac{1-f_{\min}}{\varepsilon_2} \right) \le \varepsilon_\phi \le \varepsilon_r(\text{Isotropic})
\label{eq05}
\end{equation},
where $f_{\min}=(\varepsilon_r - \varepsilon_2)/(\varepsilon_1 - \varepsilon_2)$ is the lower bound of $f$. By introducing this bound into the numerical optimization, the resulting anisotropic material can be implemented as an actual composite.

\begin{figure}[htbp]
\centering
\includegraphics[scale=1.0,clip]{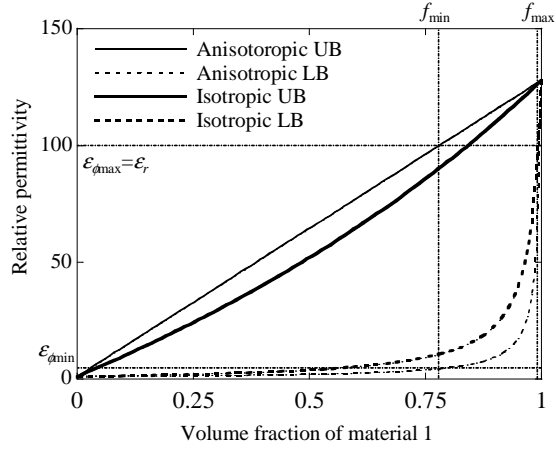}
\caption{Plot of material bounds. The thin solid and dotted lines represent anisotropic upper and lower bounds in \eqref{eq04}. The thick solid and dotted lines represent the isotropic upper and lower bounds \cite{milton2002}. The vertical and horizontal dashed lines represent the maximum and minimum values of the possible volume fraction and $\varepsilon_\phi$ for $\varepsilon_r=100$.}
\label{bound}
\end{figure}

\section{NUMERICAL RESULTS}
\subsection{Comparison of isotropic and anisotropic dielectric cloaks}
The results from the present numerical study confirm an improvement in performance as a result of introducing the anisotropic material. As a benchmark for comparison, we refer to the numerical work of Wang and Semouchkina\cite{wang2013}on isotropic dielectric cloaking. The same conditions were used in our numerical model as in their work. The frequency of the incident wave was set to 8 GHz with a wavelength $\lambda=3.75\text{cm}$. The cloaked cylindrical object had a radius of $0.75\lambda$ and the cloaked region had a thickness of $0.8695\lambda$, which was divided into eight layers. The two materials forming the composite had permittivities $\varepsilon_1=128$ and $\varepsilon_2=1$ which correspond to the permittivity limits in the optimization performed in Ref. \cite{wang2013}. The scattering width was evaluated using an integrated value over the evaluation boundary, which is represented as the total scattering width (TSCW). To clarify the cloaking effect represented by the TSCW, this was normalized with respect to the bare object's TSCW value, as reported elsewhere in the literature \cite{valagiannopoulos2012}. Finite element analysis was performed using commercial software COMSOL Multiphysics (COMSOL AB, Stockholm, Sweden). The TSCW of an optimized isotropic dielectric cloak studied in Ref. \cite{wang2013} was calculated in our system to be 0.486, which is close to the reported value of 0.502 ($=9.56/1.28$).

To better compare cloaking with isotropic and anisotropic dielectrics, isotropic dielectric cloaking was re-optimized in our system. The optimization was performed using the SNOPT algorithm \cite{gill2005} embedded in COMSOL Multiphysics. Starting from the optimized isotropic value reported in Ref. \cite{wang2013}, an optimized normalized TSCW of 0.451 was obtained, which represents a $7.4\%$ performance improvement. The resulting relative permittivities are shown in Table \ref{tableEx1} together with the results of other studies for comparison. Our result was marked by the existence of a maximum permittivity material which was not the case in Ref. \cite{wang2013} although the affordable permittivity range was set to be identical. This result was used as a reference for the performance of the optimized isotropic dielectric cloak. Optimization of the anisotropic dielectric cloak was first performed without a bound with \eqref{eq05}, that is, both $\varepsilon_r$ and $\varepsilon_\phi$ were optimized within the range $\varepsilon_2\le \varepsilon_r,\varepsilon_\phi \le \varepsilon_1$. Starting from the optimized isotropic value reported in Ref. \cite{wang2013}, an optimized and normalized TSCW of 0.387 was obtained, which represents a $14.0\%$ performance improvement compared with the optimized isotropic dielectric cloak. This result clearly shows that anisotropy of the dielectric can enhance the cloaking performance. The values in parenthesis in Table \ref{tableEx1} represent the layer number of the materials studied. Material property bounds were not satisfied in several layers, which indicate that these properties cannot be attained in a real material.

\begin{figure}[htbp]
\centering
\includegraphics[scale=1.0,clip]{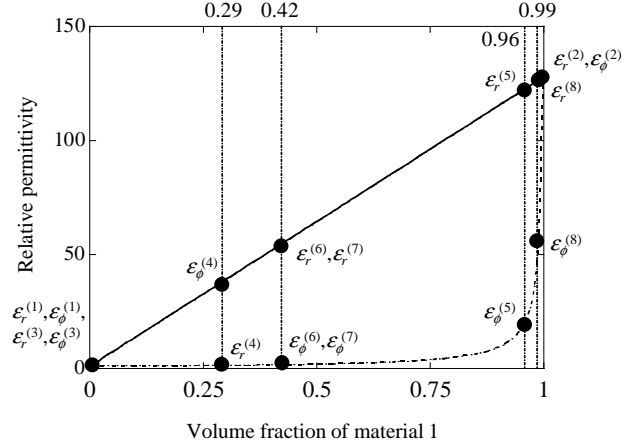}
\caption{Optimized permittivities together with the material bounds. The vertical dashed lines represent the corresponding volume fractions.}
\label{boundEx1}
\end{figure}

Optimization was subsequently performed by introducing the bound in \eqref{eq05}. To achieve a smooth optimization, we first defined which of $\varepsilon_r$ and $\varepsilon_\phi$ was larger in value. We defined these terms referring to the optimized results without material bounds and tried several patterns with the initial values obtained from the optimized results of Ref. \cite{wang2013}. The obtained optimized normalized TSCW was 0.438, which represents an approximately $2.8\%$ performance improvement. Although the percentage improvement was quite low, an improvement in performance was confirmed even in the material bound. In all isotropic layers, the values of $\varepsilon_r$ and $\varepsilon_\phi$ were at the lower and upper bounds, as shown in Fig. \ref{boundEx1}. This indicates that the maximum anisotropy within a particular volume fraction is required to improve cloaking performance. In layers 5--8, $\varepsilon_r$ was larger than $\varepsilon_\phi$, indicating that the layering direction, which is parallel to the $phi$ direction, is optimized for layers 5--8 under these conditions. These physical properties and layering patterns cannot be achieved in the multilayer isotropic dielectric.

\begin{figure}[htbp]
\centering
\includegraphics[scale=1.0,clip]{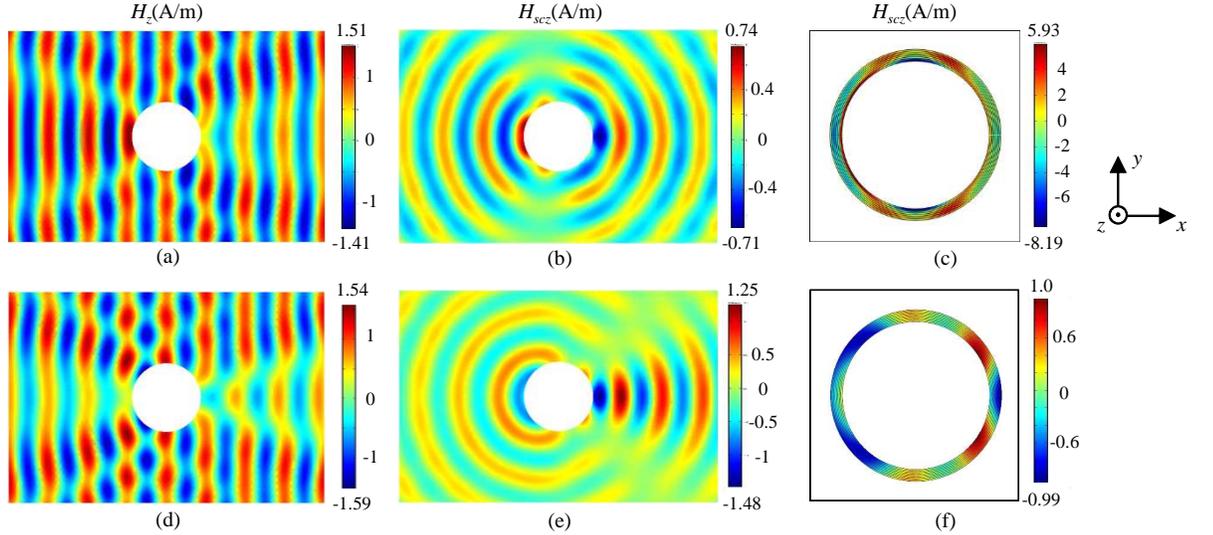}
\caption{Distributions of the total and scattered magnetic fields obtained using optimized parameters for the anisotropic dielectrics and the bare object. (a), (d): $H_z$ outside the cloak region. (b), (e): $H_{scz}$ outside the cloak region. (c), (f): $H_{scz}$ inside the cloak region. (a)--(c) represent the anisotopic dielectric cloak and (d)--(f) represent the bare object. The values on the top and bottom of the color bars represent the maximum and minimum values. }
\label{HzEx1}
\end{figure}

Magnetic field distributions are shown in Fig. \ref{HzEx1} for the optimized anisotropic dielectric and the bare object. From the total magnetic field outside the cloaked region and the bare object shown in Figs. \ref{HzEx1} (a) and (d), the reduction of reflection and screening by cloaking is confirmed. Figures \ref{HzEx1} (b), (c), (e) and (f) show the scattering of the magnetic field outside and inside the cloaked region. In the bare object, scattering clearly occurs behind the object, whereas in the anisotropic dielectric cloak, the scattering is smoothed on the right side of the cloak. Strong electric field concentration and azimuthal wave propagation were observed inside the anisotropic cloak, which reduced scattering outside the cloak.

\begin{table}[htbp]
\centering
\caption{Optimized permittivities of the eight-layer cloaking. The values in parenthesis represent the layer number of the materials.}
\label{tableEx1}%
\scriptsize
\begin{tabular}{ccccccccc}
\hline
\hline
  & $\varepsilon_{r}^{(1)}$ & $\varepsilon_{\phi}^{(1)}$ & $\varepsilon_{r}^{(2)}$ & $\varepsilon_{\phi}^{(2)}$ & $\varepsilon_{r}^{(3)}$ & $\varepsilon_{\phi}^{(3)}$ & $\varepsilon_{r}^{(4)}$ & $\varepsilon_{\phi}^{(4)}$ \\
\shortstack{Isotropic dielectrics \\ from Ref. \cite{wang2013}}& 
\multicolumn{2}{c}{1} & \multicolumn{2}{c}{65} & 
\multicolumn{2}{c}{65} & \multicolumn{2}{c}{17}\\
\shortstack{Optimal isotropic\\ dielectric}&
\multicolumn{2}{c}{1} & \multicolumn{2}{c}{128} & 
\multicolumn{2}{c}{45.48} & \multicolumn{2}{c}{1}\\
\hline
\shortstack{Anisotropic dielectrics\\ without\\ material bound}&
1 & 128 & 1 & 106 & 1 & 1 & 1 & 1 \\
\shortstack{Anisotropic dielectrics\\ with\\ material bound}&
1 & 1 & 128 & 128 & 1 & 1 & 1.41 & 38.1 \\
\hline
\hline
&$\varepsilon_{r}^{(5)}$ & $\varepsilon_{\phi}^{(5)}$ & $\varepsilon_{r}^{(6)}$ & $\varepsilon_{\phi}^{(6)}$ & $\varepsilon_{r}^{(7)}$ & $\varepsilon_{\phi}^{(7)}$ & $\varepsilon_{r}^{(8)}$ & $\varepsilon_{\phi}^{(8)}$\\
\shortstack{Isotropic dielectrics \\ from Ref. \cite{wang2013}}& 
\multicolumn{2}{c}{3} & \multicolumn{2}{c}{3} &
\multicolumn{2}{c}{4} & \multicolumn{2}{c}{58}\\
\shortstack{Optimal isotropic\\ dielectric}&
\multicolumn{2}{c}{2.88} & \multicolumn{2}{c}{9.13} & 
\multicolumn{2}{c}{10.98} & \multicolumn{2}{c}{37.15}\\
\hline
\shortstack{Anisotropic dielectrics\\ without\\ material bound}&
128 & 128 & 128 & 1.42 & 128 & 58.62 & 2.24 & 29.5\\
\shortstack{Anisotropic dielectrics\\ with\\ material bound}&
123.32 & 22.52 & 54.89 & 1.72 & 54.89 & 1.73 & 126.71 & 55.84\\
\hline
\hline
\end{tabular}
\end{table}

\subsection{Cloaking by fewer layers}
In the case that $\varepsilon_\phi>\varepsilon_r$ and both values are on the material bound, the anisotropic material represents the macroscopic state of the layered isotropic material. That is, a layer composed of the homogenous anisotropic material corresponds to several layers of the isotropic material. Accordingly, the same performance is expected for the cloak using the anisotropic dielectric as that using the isotropic dielectric consisting of a greater number of layers. Moreover, the performance difference between the isotropic and anisotropic dielectrics can be highlighted different from the eight-layer example discussed earlier.

Using the same cloaked object size ($0.75\lambda$) and cloaked region ($0.8695\lambda$) as Wang and Semouchkina \cite{wang2013}, the cloaked region was divided into three layers of the same thickness, and the optimization of the material properties was performed for both the isotropic and anisotropic materials. Since an optimized reference solution does not exist for this problem---in contrast to the previous case---the optimization was started from several patterns of random values. All other conditions were the same as in the previous case. 

\begin{figure}[htbp]
\centering
\includegraphics[scale=1.0,clip]{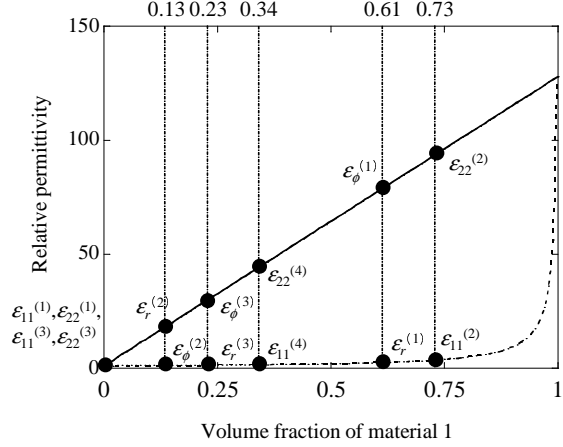}
\caption{Optimized permittivities together with the material bound. $\varepsilon_r$ and $\varepsilon_\phi$ correspond to the three-layer cylindrical cloak and $\varepsilon_{11}$ and $\varepsilon_{22}$ correspond to the octagonal cloak. The vertical dashed lines represent the corresponding volume fractions.}
\label{boundEx2}
\end{figure}

The resulting normalized TSCWs were 0.661 and 0.496 in the cloaks using isotropic and anisotropic dielectrics, respectively. A $\approx25\%$ better result was obtained using the anisotropic dielectric than the isotropic dielectric. The optimized relative permittivities are shown in Table \ref{tableEx2}. Higher permittivity is achieved in the anisotropic dielectric, which leads to its better performance. For cloaking, high permittivity in the azimuthal direction and low permittivity in the radial direction are required. This is difficult to achieve in isotropic dielectrics of only a few layers and, therefore, a lower optimized permittivity was introduced than for the anisotropic dielectric. The optimized values of the anisotropic cloak are also plotted in Fig. \ref{boundEx2} together with the curve of the material bound. In all anisotropic layers, values of $\varepsilon_r$ and $\varepsilon_\phi$ are on the lower and upper bound.

\begin{figure}[htbp]
\centering
\includegraphics[scale=1.0,clip]{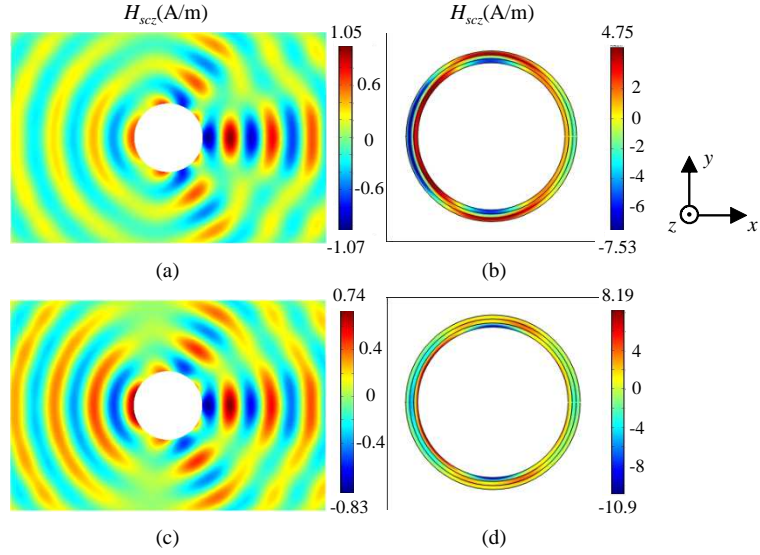}
\caption{Distributions of the scattered magnetic field using optimized parameters with isotropic and anisotropic dielectrics. (a), (c): $H_{scz}$ outside the cloak. (b), (d): $H_{scz}$ inside the cloak. (a) and (b) represent the isotopic dielectric cloak and (c) and (d) represent the anisotropic dielectric cloak. The values on the top and bottom of the color bars represent the maximum and minimum values. }
\label{HzEx2}
\end{figure}

\begin{figure}[htbp]
\centering
\includegraphics[scale=1.0,clip]{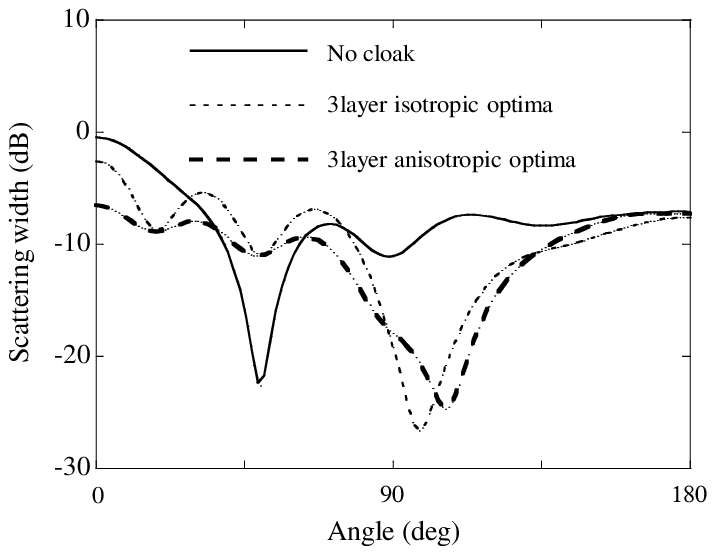}
\caption{Scattering widths of the bare object and cloaks with isotropic and anisotropic dielectrics.}
\label{swEx2}
\end{figure}

Figure \ref{HzEx2} shows the scattering of the magnetic field outside and inside both the isotropic and anisotropic cloaks. The distributions of the scattering width of the bare object, the three-layer isotopic cloak, and the anisotropic cloak are shown in Figure \ref{swEx2}. At an angle of $0^\circ$, strong scattering was observed in the isotropic cloak, indicating that the wave did not bend sufficiently to the right side of the object. The maximum value of the range between the maximum and minimum scattering values is greater inside the anisotropic cloak than inside the isotropic cloak. Owing to the strong anisotropy inside the cloak, strong magnetic field concentration results, which reduces scattering outside the cloak.

The optimized TSCW value of 0.496 is close to that published for an eight-layer isotropic cloak in Ref. \cite{wang2013} (0.486). This indicates that an acceptable cloaking performance can be achieved by using just a few anisotropic dielectric layers.

\begin{table}[htbp]
\caption{Optimized permittivities for three-layer cloaking. The values in parenthesis represent the layer number of the materials.}
\label{tableEx2}%
\begin{tabular}{ccccccc}
\hline
\hline
& $\varepsilon_r^{(1)}$ & $\varepsilon_\phi^{(1)}$ & $\varepsilon_r^{(2)}$ & $\varepsilon_\phi^{(2)}$ & $\varepsilon_r^{(3)}$ & $\varepsilon_\phi^{(3)}$\\
\hline
Isotropic dielectric&
\multicolumn{2}{c}{1} & \multicolumn{2}{c}{49.68} &
\multicolumn{2}{c}{73.43}\\
Anisotropic dielectric&
2.55 & 78.9 & 17.18 & 1.14 & 1.30 & 30.76\\
\hline
\hline
\end{tabular}
\end{table}

\subsection{Cloaking of a non-circular object}

\begin{figure}[htbp]
\centering
\includegraphics[scale=1.0,clip]{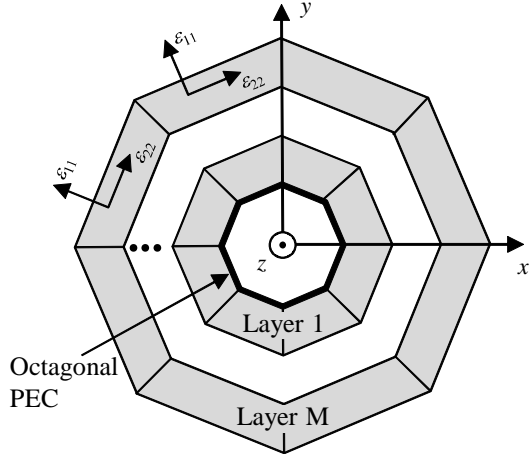}
\caption{Outline of the multilayer dielectric cloaking for an octagon PEC. The principal directions of the anisotropic material are set parallel and perpendicular to each line.}
\label{octagon}
\end{figure}

Because this study is based on an optimization methodology using full-wave finite element analysis, a non-circular cloaking object can also be handled. An octagon cloaking for an octagon PEC was designed, as shown in Fig. \ref{octagon}, to confirm the effectiveness of the anisotropic material for cloaking a non-circular shape. As with cylindrical cloaking, each layer was composed of the same homogenous anisotropic material. The octagon was divided into eight parts in each layer according to the line layout. The principal directions of the composite were fixed in each part and set perpendicular and parallel to each line. The number of layers was set to 4. The octagonal shape did not have circular symmetry and the cloaking effect varied within each $22.5^\circ$ incident angle. To achieve a constant cloaking effect in these angles, the optimizations were performed for two incident waves from different angles ($0^\circ$ and $22.5^\circ$). Two TSCWs obtained by each analysis were simultaneously minimized by forming the objective function by adding these values.

\begin{figure}[htbp]
\centering
\includegraphics[scale=1.0,clip]{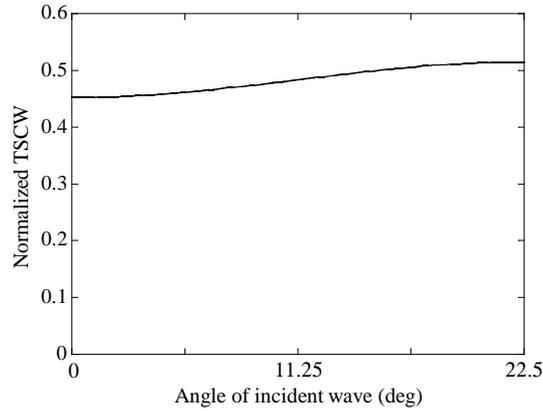}
\caption{Dependence of TSCW on the incident wave angle.}
\label{angleEx3}
\end{figure}

The resulting TSCWs were 0.453 and 0.513 for $0^\circ$ and $22.5^\circ$ incident waves. Table \ref{tableEx3} shows optimized relative permittivity values, which are plotted in Fig. \ref{boundEx2} together with the previous optimized results. As with the circular object, the optimized dielectric values of anisotropic layers are located on the material physical property bounds. Thus, the large anisotropy can be effective even in non-circular cloaking. Figure \ref{angleEx3} shows the total magnetic field distributions of cloaked and bare objects for $0^\circ$ and $22.5^\circ$ incident waves. The reflections and shadows were reduced in the cloaked object. In particular, for the $22.5^\circ$ direction incident wave, a clear shadow was observed behind the bare object. Since the wide, straight line is perpendicular to the wave direction, it is difficult for the wave to go around the object compared with the $0^\circ$ incident wave. Nevertheless, cloaking was effective even for such an object. The TSCWs analyzed in the angles between $0^\circ$ and $22.5^\circ$ are plotted in Fig. \ref{angleEx3}. Smooth interpolation was obtained between these angles and the resulting incident angle with the minimum cloaking effect was $22.5^\circ$. That is, octagon cloaking has a cloaking effect with a TSCW between 0.453 and 0.513 from the incident wave from any direction.

\begin{table}[htbp]
\caption{Optimized values of anisotropic permittivities of four-layer octagonal cloaking. The values in parenthesis represent the layer number of the materials.}
\label{tableEx3}%
\begin{tabular}{ccccccccc}
\hline
\hline
& $\varepsilon_{11}^{(1)}$ & $\varepsilon_{22}^{(1)}$ & $\varepsilon_{11}^{(2)}$ & $\varepsilon_{22}^{(2)}$ & $\varepsilon_{11}^{(3)}$ & $\varepsilon_{22}^{(3)}$ & $\varepsilon_{11}^{(4)}$ & $\varepsilon_{22}^{(4)}$\\
\hline
Anisotropic dielectrics&
1 & 1 & 3.62 & 93.65 & 1 & 1 & 1.5 & 43.69\\
\hline
\hline
\end{tabular}
\end{table}

\begin{figure}[htbp]
\centering
\includegraphics[scale=1.0,clip]{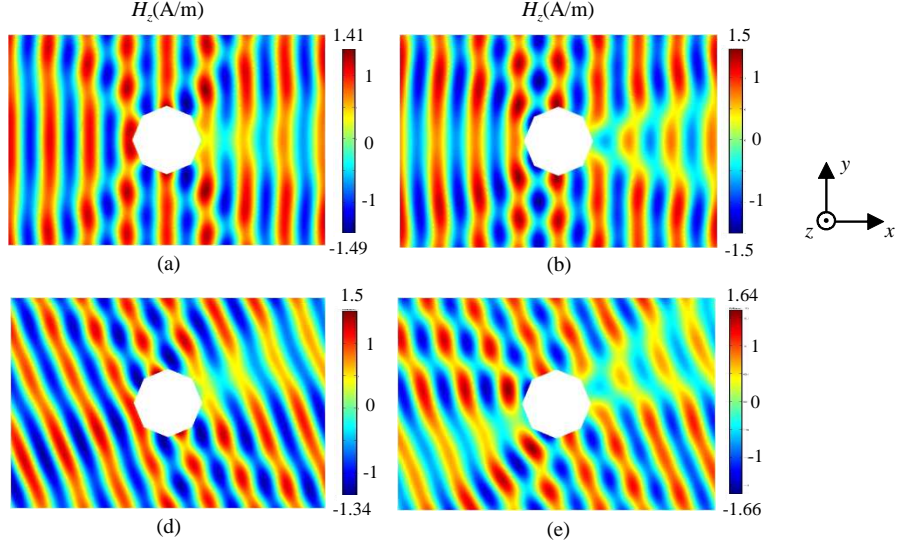}
\caption{Distributions of the total magnetic field obtained using optimized parameters for a bare object with $0^\circ$ and $22.5^\circ$ incident waves. (a), (c): $H_z$ outside the cloak with cloaking. (b), (d): $H_z$ outside the cloaking region without cloaking. (a) and (b) are the results of $0^\circ$ incident wave and (c) and (d) are the results of $22.5^\circ$ incident wave. The values on the top and bottom of the color bars represent the maximum and minimum values.}
\label{HzEx3}
\end{figure}

\section{CONCLUSION}
We studied and identified the effectiveness of an anisotropic dielectric composite for non-resonant cloaking using numerical optimization techniques based on full-wave finite element analysis. Because the anisotropic material properties can be regarded as macroscopic properties of a layered isotropic dielectric, one layer composed of anisotropic material worked as effectively as a multilayer isotropic material.

For eight- and three-layer cylindrical cloaking, the cloaking performance was about $2.8\%$ and $25.0\%$ higher, respectively, when the anisotropic dielectric was introduced compared with the isotropic material. The performance of three-layer cloaking using the anisotropic dielectric was similar to that published for eight-layer cloaking using an isotropic dielectric.\cite{wang2013} All microstructures based on the optimized values of anisotropic permittivities were identified as simple laminations, which correspond to the Wiener upper and lower bounds. We identified that approximately 50\% cloaking---according to the evaluation criteria of scattering width---can be achieved using just a few layers of the laminated composite. In practice, if a composite with the desired physical properties could be fabricated, cloaking could be achieved by manufacturing just a few layers. When considering the total manufacturing cost of the device and the composite, it is unclear how this would compare with existing methods. However, such an anisotropic dielectric cloaking material could be easily implemented in bulk production, that is, when many devices are manufactured from the material block.

Because our study was based on a full-wave simulation, anisotropic dielectric cloaks could be designed for arbitrary shapes, as exemplified in this paper by an octagon. Its flexibility also allows this methodology to be extended to various devices proposed in the context of transformation optics \cite{chen2010nmat,liu2012} by changing the evaluation criteria of the devices.

\end{document}